\documentclass[prd,aps,preprint,tightenlines,showpacs,nofootinbib,superscriptaddress]{revtex4-1}
\usepackage{mathrsfs}
\usepackage{amsfonts}
\usepackage{amsmath}
\usepackage{array}
\usepackage{verbatim}
\usepackage{bm}
\usepackage{epsfig}
\usepackage{relsize}
\usepackage{lineno}
\usepackage{float}
\usepackage{multirow}
\RequirePackage{xspace}

\def\lsim{\raise0.3ex\hbox{$<$\kern-0.75em\raise-1.1ex\hbox{$\sim$}}}
\def\gsim{\raise0.3ex\hbox{$>$\kern-0.75em\raise-1.1ex\hbox{$\sim$}}}

\def\beq{\begin{equation}}
\def\eeq{\end{equation}}
\def\bea{\begin{eqnarray}}
\def\eea{\end{eqnarray}}
\def\bq{\begin{quote}}
\def\eq{\end{quote}}

\def\gappeq{\mathrel{\rlap {\raise.5ex\hbox{$>$}}
{\lower.5ex\hbox{$\sim$}}}}

\def\lappeq{\mathrel{\rlap{\raise.5ex\hbox{$<$}}
{\lower.5ex\hbox{$\sim$}}}}

\def\Toprel#1\over#2{\mathrel{\mathop{#2}\limits^{#1}}}

\begin{document}


\title{Impact of intrinsic charm amount in the nucleon 
and saturation effects on 
the prompt atmospheric \bm{$\nu_{\mu}$} flux for IceCube}

\author{Victor P.~Goncalves}
\email{barros@ufpel.edu.br}
\affiliation{Instituto de F\'{\i}sica e Matem\'atica,  Universidade
Federal de Pelotas (UFPel), \\
Caixa Postal 354, CEP 96010-900, Pelotas, RS, Brazil}

\author{Rafa{\l} Maciu{\l}a}
\email{rafal.maciula@ifj.edu.pl} \affiliation{Institute of Nuclear
Physics, Polish Academy of Sciences, Radzikowskiego 152, PL-31-342 Krak{\'o}w, Poland}

\author{Antoni Szczurek\footnote{also at University of Rzesz\'ow, PL-35-959 Rzesz\'ow, Poland}}
\email{antoni.szczurek@ifj.edu.pl} \affiliation{Institute of Nuclear
Physics, Polish Academy of Sciences, Radzikowskiego 152, PL-31-342 Krak{\'o}w, Poland}

\date{\today}

\begin{abstract}
The predictions for the atmospheric neutrino flux at high energies
strongly depend on the contribution of prompt neutrinos, which are
determined by the production of charmed meson in the atmosphere 
at very forward rapidities. In this paper we estimate the related 
cross sections
taking into account the presence of an intrinsic charm (IC) component 
in the proton wave function and the QCD dynamics modified by the onset of
saturation effects. 
The impact on the predictions for the prompt neutrino flux is
investigated assuming different values for the probability to find the
IC in the nucleon.
Comparison with the IceCube data is performed and conclusions are drawn.

\end{abstract}

\maketitle

\section{Introduction}
\label{intro}

The understanding of Particle Physics have been challenged and improved by the recent experimental results obtained by the LHC, the Pierre
Auger and  IceCube Neutrino Observatories \cite{pdg}. In particular, in
recent years, IceCube measured the astrophysical and atmospheric
neutrinos fluxes at high energies
\cite{ice1,Aartsen:2016xlq,Aartsen:2017nbu} and different collaborations
from the LHC performed several analyses of the heavy meson production at
high energies and forward rapidities
\cite{Aaij:2013mga,Aaij:2015bpa,Acharya:2017jgo,Acharya:2019mgn}.  
Such distinct sets of data are intrinsically related, since the
description of the heavy meson production  
at the LHC and higher center of mass energies is fundamental to make
predictions of the prompt neutrino flux \cite{Goncalves:2017lvq}, which
is expected to dominate the atmospheric  $\nu$ flux for large neutrino energies
\cite{review_neutrinos}. An important question, 
which motivate the present study, is whether the current and future
IceCube data can shed light on charm production at the LHC and vice -
versa and in particular on the intrinsic charm in the nucleon.

\begin{figure}[t]
\includegraphics[scale=0.5]{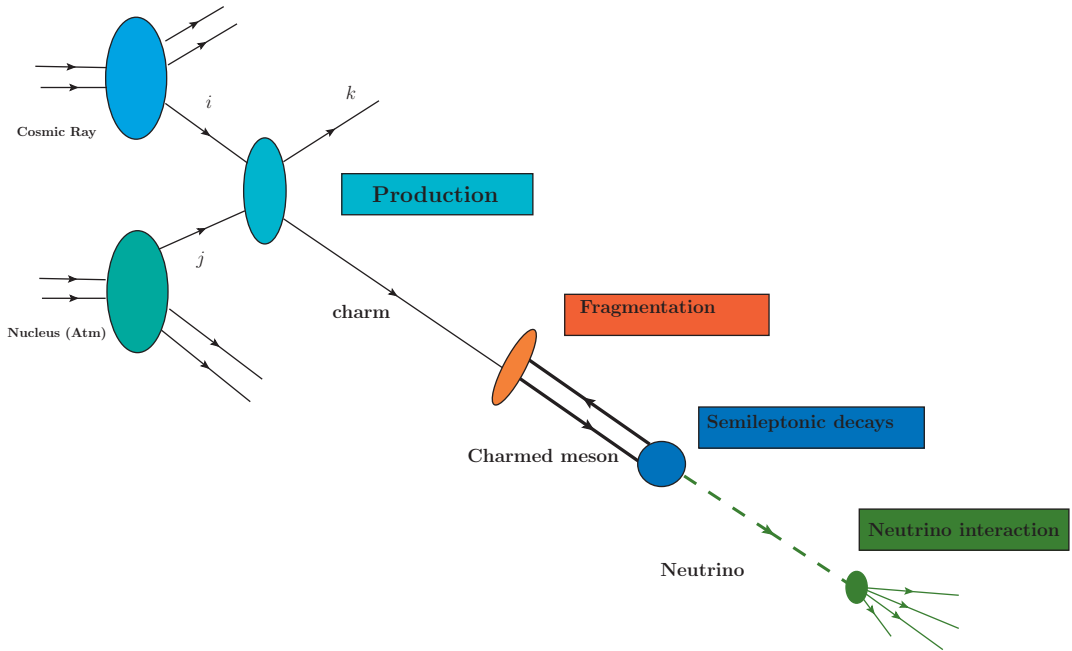}
\caption{Representation of the ingredients needed to estimate the prompt neutrino flux at the detector level. }
\label{Fig:diagrama}
\end{figure}

In order to derive  realistic predictions of the prompt atmospheric neutrino flux at the
detector level  we should have theoretical control of the description of
several ingredients (see Fig.~\ref{Fig:diagrama}): the incident cosmic
flux, the charm production, its hadronization, the decay of the
heavy hadrons,  the propagation of the associated particles
through the atmosphere and the neutrino interaction (see e.g. 
Refs.~\cite{sigl,anna1,gauld,halzen,anna2,laha,prosa,prdandre,Goncalves:2018zzf}).
As demonstrated in our previous study \cite{Goncalves:2017lvq},   to address the production of 
high-energy neutrinos ($E_{\nu} >$ 10$^5$ GeV), it is fundamental to
precisely describe  the charmed meson production at very high energies
and large forward rapidities.  This aspect motivated the development
of new and/or more precise approaches to describe the perturbative and
nonperturbative regimes of the Quantum Chromodynamics (QCD) needed to
describe the charmed meson production in a kinematical range  beyond
that reached in hadronic collisions at the LHC. For this new kinematical
range, some topics are theme of intense debate: (a) the presence (or
not) of intrinsic heavy quarks in the hadronic wave function
\cite{bhps,pnndb,wally99}, characterized by a large value of 
the longitudinal momentum fraction of beam nucleon momentum; 
(b) the validity of the collinear factorization at high energies
\cite{nikwolf,raju,Dominguez:2010xd,Kotko:2015ura}, since it disregards
the transverse momentum of the incident particles; and (c) the presence
(or not) of nonlinear (saturation) effects on the description of the QCD
dynamics at high energies \cite{cgc}, which are expected to contribute
at high energies due to the high partonic density predicted by  linear
DGLAP or BFKL evolution equations; 
(d) the impact of subleading fragmentation of light partons
on heavy meson production at high energies and very forward rapidities
and its consequences for prompt neutrino flux \cite{Maciula:2017wov,Goncalves:2018zzf}.
Such questions naturally arise due to the fact that 
in the calculation of the prompt neutrino flux at high energies, the
main contribution for the charm production cross section comes from
partons with very small (large) values of  $x$  in the hadrons that
constitute the atmosphere (incident cosmic ray flux). Recently, two of
us have presented in Ref.~\cite{antoni_rafal_jhep} a comprehensive study
of the charm production at large rapidities considering the collinear,
hybrid and $k_T$-factorization approaches taking into account
the presence of an intrinsic charm in the proton wave function with
parton distributions that are solutions of linear and nonlinear
evolution equations. One of the goals of this paper is to extend the
analysis performed in Ref.~\cite{antoni_rafal_jhep} and derive
associated prompt neutrino fluxes at high energies. In particular, we
shall estimate the impact of the intrinsic charm -- initiated subprocess
and/or saturation effects on the predictions for the prompt neutrino flux. 
Another more ambitious goal is to verify whether the recent IceCube data
for the prompt $\nu_{\mu}$ flux allow to derive an upper bound for the probability 
of finding a charm quark-antiquark pair in the proton wave function, 
which is one of the main uncertainties in the modelling of 
the intrinsic charm. A similar goal was also present in the 
analyses performed in Refs. \cite{halzen,laha}. However, our study
differs from these previous analyses in several aspects. 
Our predictions for the $x_F$ distributions will be derived using a
framework that sucessfully describes the LHC data, with the main input
being the parton distribution functions which were derived using the
world data. 
In these previous studies, the $x_F$ distribution was fitted using the
old data for the $D$ and $\Lambda_c$ production, with the normalization 
being a parameter free. 
Moreover, the energy dependence of the intrinsic charm contribution was
assumed to follow the inelastic cross section, which is dicted by soft
processes. In contrast, in our approach, such contribution is 
calculated perturbatively, which implies a steeper energy dependence. 
Finally, our predictions are calculated using a unified approach for 
the $gg \rightarrow c \bar{c}$ and $gc \rightarrow gc$ mechanisms, 
which we believe to be more realistic to treat the charm production 
at central and forward rapidities.

The paper is organized as follows. In the next section a brief review of
formalism needed to estimate the prompt $\nu_{\mu}$ flux 
is presented. In particular,
we discuss the  $Z$-moment method \cite{ingelman}, the hybrid approach
for production of $c/\bar c$ quarks/antiquarks
and the main inputs and underlying assumptions of our calculations. In
Section \ref{res}, we shall present our predictions for the Feynman $x_F$
distribution and for the prompt flux considering different charm
production mechanisms and different models for the unintegrated gluon
distribution. Moreover, the prompt flux is estimated assuming different
amounts for the probability of finding an intrinsic charm
component in the nucleon and the predictions are compared with 
the recent IceCube data. 
Finally, in Section \ref{conc} we shall summarize our main results 
and formulate conclusions.

\begin{figure}[t]
\includegraphics[scale=0.2]{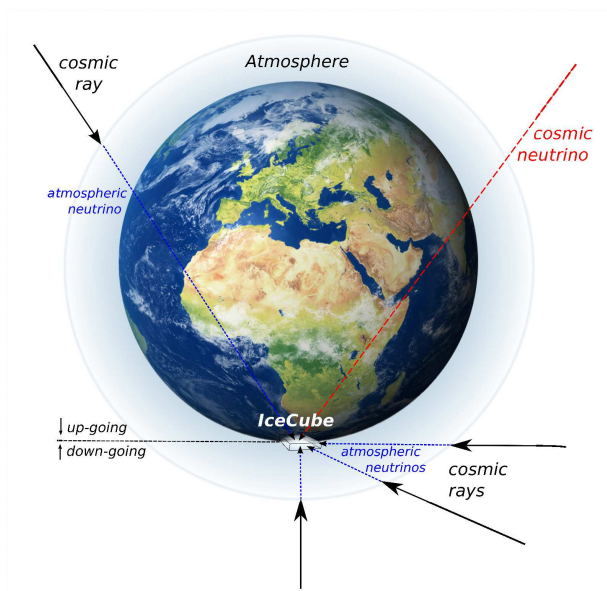}
\caption{A schematic illustration of the IceCube experiment.}
\label{Fig:IceCube}
\end{figure} 

\section{Formalism}

A schematic illustration of the IceCube experiment is shown in Fig.~\ref{Fig:IceCube}. Neutrinos are detected through the
Cherenkov light emitted by secondary particles produced
in neutrino-nucleon interactions in or around the detector. Although primarily designed for the detection of high-energy
neutrinos from astrophysical sources, denoted cosmic neutrino in Fig.~\ref{Fig:IceCube},  IceCube can also be
used for investigating the atmospheric neutrino spectrum.  
The atmospheric neutrinos are produced in cosmic-ray interactions with nuclei
in Earth's atmosphere \cite{review_neutrinos}. While at low neutrino
energies ($E_{\nu}\lesssim 10^5$ GeV), these neutrinos arise from the
decay of light mesons (pions and kaons), and the associated flux is
denoted as the {\it conventional} atmospheric neutrino flux 
\cite{Honda:2006qj}, for larger energies  
it is expected that the {\it prompt} atmospheric neutrino flux associated with the
decay of hadrons containing heavy flavours become important
\cite{ingelman}. 
One has that the flux of conventional atmospheric neutrinos is a function
of the zenith angle, since horizontally travelling mesons have
a much higher probability to decay before losing energy in
collisions, which implies  a harder conventional neutrino spectrum of
horizontal events compared to vertical events. In contrast, heavy mesons decay before interacting and follow the initial
spectrum of cosmic rays more closely, being almost independent of the zenith angle in the neutrino energy range probed by the IceCube.
As discussed in the Introduction, 
the calculation of the prompt atmospheric neutrino flux at the
detector level depends on the description of the  production and decay of the
heavy hadrons as well as the propagation of the associated particles
through the atmosphere (see Fig.~\ref{Fig:diagrama}). Following our
previous studies \cite{Goncalves:2017lvq,Goncalves:2018zzf}, we will
estimate the expected prompt neutrino flux in the detector 
$\phi_{\nu}$  using the $Z$-moment method \cite{ingelman}, which implies
that  $\phi_{\nu}$  can be estimated using the geometric interpolation formula
\begin{eqnarray}
\phi_{\nu} = \sum_H \frac{\phi_{\nu}^{H,low} \cdot  \phi_{\nu}^{H,high}} {\phi_{\nu}^{H,low} + \phi_{\nu}^{H,high}} \,\,.
\label{eq:flux}
\end{eqnarray}
where $H = D^0, D^+, D_s^+$, $\Lambda_c$ for charmed hadrons and
${\phi_{\nu}^{H,low}}$ and ${\phi_{\nu}^{H,high}}$ are solutions of a
set of coupled cascade equations for the nucleons, heavy mesons 
and leptons (and their antiparticles) 
fluxes in the low- and high-energy ranges, respectively. 
They can be expressed in terms of the nucleon-to-hadron
($Z_{NH}$), nucleon-to-nucleon ($Z_{NN}$), hadron-to-hadron  ($Z_{HH}$) 
and hadron-to-neutrino ($Z_{H\nu}$) $Z$-moments,  as follows \cite{ingelman}
\begin{eqnarray}
\phi_{\nu}^{H,low} & = & \frac{Z_{NH}(E) \, Z_{H\nu}(E)}{1 - Z_{NN}(E)} \phi_N (E,0) \,, \\
\phi_{\nu}^{H,high} & = & \frac{Z_{NH}(E) \, Z_{H\nu}(E)}{1 - Z_{NN}(E)}\frac{\ln(\Lambda_H/\Lambda_N)}
{1 - \Lambda_N/\Lambda_H} \frac{m_H c h_0}{E \tau_H} f(\theta) \, \phi_N (E,0) \,,
\end{eqnarray}
where $\phi_N(E,0)$ is the primary 
flux of nucleons in the atmosphere, $m_H$ is the decaying particle's mass, $\tau_H$ is the proper 
lifetime of the hadron, $h_0 = 6.4$ km, 
$f(\theta) \approx 1/\cos \theta$ for $\theta < 60^o$, and the effective 
interaction lengths $\Lambda_i$ are given by $\Lambda_i = \lambda_i/(1 -
Z_{ii})$, with $\lambda_i$ being the associated interaction length ($i =
N,H$). For $Z_{H\nu}$, our treatment of the semileptonic decay of $D$-hadrons follows closely Ref. \cite{anna2}. In particular, we assume the analytical decay distributions $H \rightarrow \mu \nu_{\mu}X$ obtained in Ref. \cite{Bugaev:1998bi} and use  the decay branching ratios reported in the most recent PDG \cite{pdg}.
For a detailed discussion of the cascade equations, see
e.g.~Refs.~\cite{ingelman,sigl}. 
Assuming that the incident flux can be represented by protons ($N = p$), the charmed 
hadron $Z$-moments are given by
\begin{eqnarray}
Z_{pH} (E) =  \int_0^1 \frac{dx_F}{x_F} \frac{\phi_p(E/x_F)}{\phi_p(E)} 
\frac{1}{\sigma_{pA}(E)} \frac{d\sigma_{pA \rightarrow H}(E/x_F)}{dx_F} \,\,,
\label{eq:zpc}
\end{eqnarray}
where $E$ is the energy of the produced particle (charmed meson), $x_F$ 
is the Feynman variable, $\sigma_{pA}$ is the inelastic proton-Air
cross section and 
$d\sigma/dx_F$ is the differential cross section for the charmed meson 
production. Following previous 
studies \cite{sigl,anna1,gauld,halzen,anna2,laha,prosa,prdandre,Goncalves:2018zzf},
we will assume that $A = 14$, i.e. we will
take the $^{14}N$ nucleus as the most representative element in 
the composition of the atmosphere. 
For this value of the atomic mass number, it is a reasonable
approximation to assume that $\sigma_{pA \rightarrow charm}  \approx
A \times \sigma_{pp \rightarrow charm}$. Surely a more refine analysis
of these two aspects is possible but would shadow our discussion
of the selected issues. For $\sigma_{pA}$ we will assume the prediction
presented in Ref.~\cite{Ostapchenko:2005nj} (for a more detailed
discussion see Ref.~\cite{vicandre}).

The transition from quarks to hadrons in our calculations is done within
the independent parton fragmentation picture 
(see e.g. Ref.~\cite{Maciula:2019iak}).
It is done assuming that the hadron pseudorapidity is equal to parton 
pseudorapidity and only momenta of hadrons are reduced compared to 
the parent partons. In such an approximation the charmed meson  
$x_F$-distributions at large $x_F$ can be obtained from 
the charm quark/antiquark $x^{c}_{F}$-distributions as:
\begin{eqnarray}
\frac{d \sigma_{pp \rightarrow H} (x_{F})}{d x_{F}} =  \int_{x_{F}}^{1} \frac{dz}{z} \frac{d \sigma_{pp \rightarrow charm} (x^{c}_{F})}{d x^{c}_{F}} D_{c\to H}(z),
\label{eq:dif3}
\end{eqnarray}
where $x^{c}_{F} = x_{F}/z$ and $D_{c\to H}(z)$ is the relevant 
fragmentation function (FF). Here, in the numerical calculations we take
the traditional Peterson FF \cite{Peterson:1982ak} with 
$\varepsilon = 0.05$. The resulting meson distributions are 
further normalized by the proper fragmentation probabilities.

\begin{figure}[t]
\begin{center}
\begin{tabular}{cc}
\includegraphics[width=0.4\textwidth]{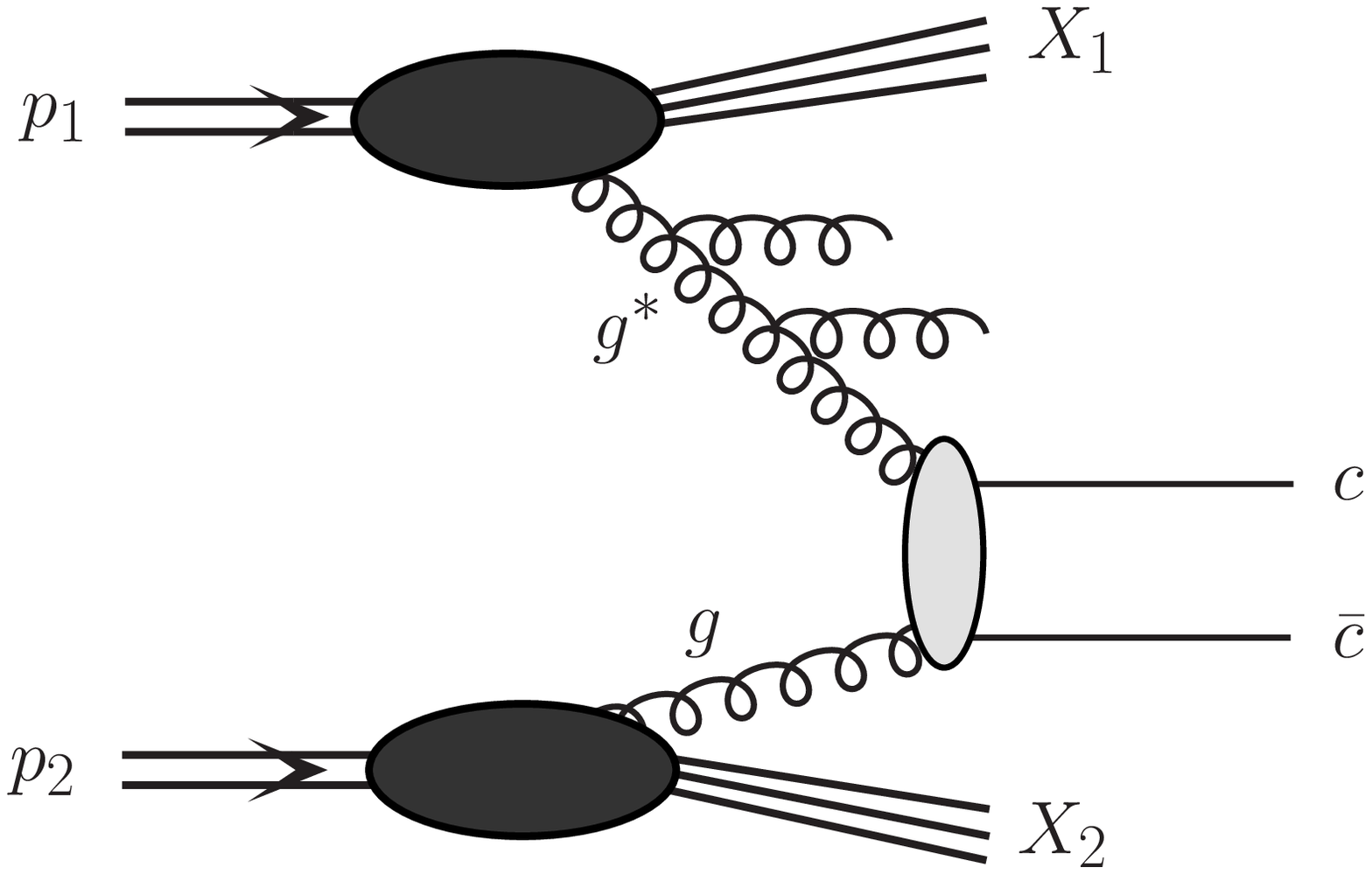} &  
\includegraphics[width=0.4\textwidth]{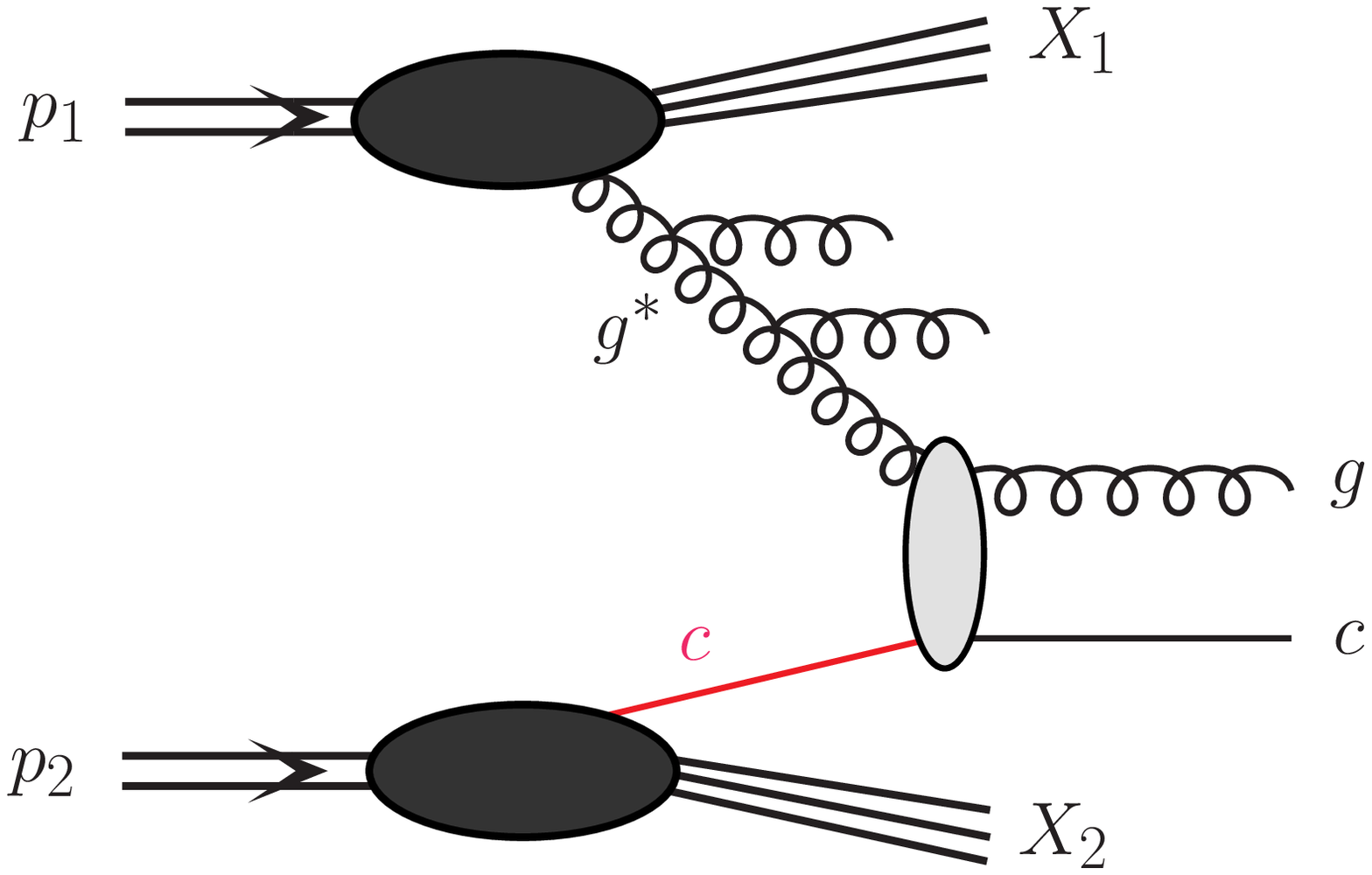} \\
(a) & (b)
\end{tabular}

\end{center}
\caption{A sketch of the (a) $gg^* \rightarrow  c\bar{c}$ and (b) $cg^* \rightarrow  cg$  production mechanisms in $pp$-interactions within the hybrid model.}
\label{Fig:diagrams}
\end{figure} 

As discussed in Ref.~\cite{antoni_rafal_jhep}, the cross section for the
charm production at large forward rapidities, which is the region of
interest for estimating the prompt $\nu_{\mu}$ flux
\cite{Goncalves:2017lvq}, can be expressed as follows
\begin{eqnarray}
d \sigma_{pp \rightarrow charm} = d \sigma_{pp \rightarrow charm}(gg \rightarrow c \bar{c}) + d \sigma_{pp \rightarrow charm}(cg \rightarrow  cg) \,\,,
\label{eq:fac}
\end{eqnarray}
where the first and second terms represent the contributions associated
with the  $gg \rightarrow c \bar{c}$ and $ cg \rightarrow cg $ mechanisms,
with the  corresponding expressions depending  on the factorization
scheme assumed in the calculations. In
Ref.~\cite{antoni_rafal_jhep}, a detailed comparison between the
collinear, hybrid and $k_T$-factorization approaches was performed. 
In what follows, we will focus  on the hybrid factorization model, which
is based on the studies performed also in 
Refs.~\cite{nikwolf,raju,Dominguez:2010xd,Kotko:2015ura}. 
Such a choice is motivated  by: 
(a) the theoretical expectation that the
collinear approach, largely used in previous calculations of
$\phi_{\nu}$, breaks down at very small-$x$ \cite{raju,Kotko:2015ura};
and that (b) the $k_T$-factorization approach reduces to the hybrid
model in the dilute-dense regime, which is the case in the charm
production at very forward rapidities, where we are probing large
(small) values of $x$ in the projectile (target). 
In this approach, the differential cross sections for $gg^* \rightarrow
c\bar{c}$ and $cg^* \rightarrow  cg$  mechanisms, 
sketched in Fig.~\ref{Fig:diagrams}, are given by
\begin{eqnarray}
d \sigma_{pp \rightarrow charm}(gg \rightarrow c \bar{c}) = \int dx_1  \int \frac{dx_2}{x_2} \int d^2k_t \, g(x_1,\mu^2) \, {\cal{F}}_{g^*} (x_2, k_t^2, \mu^2) \, d\hat{\sigma}_{gg^* \rightarrow  c\bar{c}} 
\label{eq:dif}
\end{eqnarray}
and
\begin{eqnarray}
d \sigma_{pp \rightarrow charm}(cg \rightarrow  cg) = \int dx_1  \int \frac{dx_2}{x_2} \int d^2k_t \, c(x_1,\mu^2) \, {\cal{F}}_{g^*} (x_2, k_t^2, \mu^2) \, d\hat{\sigma}_{cg^* \rightarrow  cg} \,\,,
\label{eq:dif2}
\end{eqnarray}
where $g(x_1,\mu^2)$ and $c(x_1,\mu^2)$ are the collinear PDFs in the
projectile, ${\cal{F}}_{g^*} (x_2, k_t^2, \mu^2)$ is the unintegrated
gluon distribution (gluon uPDF) of the proton target, $\mu^2$ 
is the factorization scale of the hard process and the subprocesses 
cross sections are calculated assuming that the small-$x$ gluon is off 
mass shell and are obtained from a gauge invariant tree-level off-shell 
amplitude. In our calculations $c(x_1,\mu^2)$, similarly 
$\bar c(x_1,\mu^2)$, contain the intrinsic charm component.

As emphasized in Ref.~\cite{antoni_rafal_jhep}, the hybrid 
model, already at leading-order,
takes into account radiative higher-order corrections associated with
extra hard emissions that are resummed by the gluon uPDF.
In the numerical calculations below the intrinsic charm PDFs are taken at the initial scale $m_c=  1.3$ GeV,  so the perturbative charm contribution is intentionally  not  taken into account when discussing IC contributions.

Considering the $cg^* \rightarrow  cg$ mechanism one has to deal with
the massless partons (minijets) in the  final state. The relevant
formalism with massive partons is not yet available. Therefore it is
necessary to regularize the cross section that has a singularity in the
$p_{t} \rightarrow 0 $ limit. We follow here the known prescription
adopted in \textsc{Pythia}, where a special suppression factor is
introduced at the cross section level. The form factor depends on a free
parameter $p_{t0}$, which will be fixed here using experimental
  data for the $D$ meson production in $p + p$ and $p+^4He$ collisions at 
$\sqrt{s} = 38.7$ GeV and 86 GeV, respectively.

The predictions for the charm production strongly depend on the
modelling of the partonic content of the proton
~\cite{antoni_rafal_jhep}. In particular, the contribution of the charm
- initiated process is directly associated with the description of the
extrinsic and intrinsic components of the charm content in the proton
(for a recent review see, e.g. Ref.~\cite{Brodsky:2020zdq}).  
Differently from the extrinsic charm quarks/antiquarks that 
are generated perturbatively by gluon splitting, the intrinsic 
one have multiple connections to  the valence quarks of the proton 
and thus is sensitive to its nonperturbative structure
\cite{bhps,pnndb,wally99}. 
The presence of an intrinsic component implies a large enhancement 
of the charm distribution  at large  $x$  ($> 0.1$) in comparison to 
the extrinsic charm prediction. Moreover, due to the momentum sum rule, 
the gluon distribution is also modified by the inclusion of intrinsic
charm. In recent years, 
the presence of an intrinsic charm (IC) component  have been included in
the initial conditions of the global parton analysis \cite{nnpdf,ct14}, 
the resulting IC distributions that are compatible with the world
experimental data. 
However, its  existence is still a subject of intense debate
\cite{wally16,brod_letter}, mainly associated with the amount of intrinsic
charm in the  proton wave function, which is directly related to 
the magnitude of the probability to find an intrinsic charm or anticharm
($P_{ic}$) in the nucleon.

In our analysis we will consider the collinear PDFs
given by the CT14nnloIC parametrization \cite{ct14} from
a global analysis assuming that the $x$-dependence of
the intrinsic charm component is described by the BHPS model
\cite{bhps}. In this model the proton light cone wave function has
higher Fock states, one of them being $|q q q c \overline{c}>$. 
The cross sections will be initially estimated in the next section using
the set obtained for  $P_{ic} = 1\%$ and, for comparison, the results
for the case without IC will also be presented. Another important
ingredient is the modelling of ${\cal{F}}_{g^*} (x_2, k_t^2, \mu^2)$,
which depends on the treatment of the QCD dynamics 
for the unintegrated gluon distribution at small-$x$. 
Currently, there are several models in the literature, some of them have
been reviewed in Ref.~\cite{antoni_rafal_jhep}. In our analysis 
we shall consider three different models: 
two based on the solutions of linear evolution equations, which
disregard nonlinear (saturation effects) and one being the solution of
the Balitsky-Kovchegov equation \cite{bk}, which takes into account
these effects in the small-$x$ regime. 
In particular, we will use the uPDF derived using the Kimber-Martin-Ryskin (KMR) prescription \cite{kmr}, which assumes that the transverse
momentum of the partons along the evolution ladder is strongly ordered
up to the final evolution step. In the last step this
assumption breaks down and the incoming parton that enters into 
the hard interaction posses a large transverse momentum ($k_t \approx
\mu$). Such prescription allow us to express ${\cal{F}}_{g^*} (x_2,
k_t^2, \mu^2)$ in terms of Sudakov form factor, which resums all the
virtual contributions from the scale $k_t$ to the scale $\mu$, and a
collinear $g$ PDF, which satisfies the DGLAP evolution equations. 
For this model, we will estimate the  uPDF using as input the CT14nnlo
parametrization (with and without IC) \cite{ct14} and the associated
predictions will be denoted as KMR hereafter.
Some time ago we showed that in the case of charm production at the LHC,
the KMR uPDF leads to a reasonable description of the experimental data for
$D$-meson and $D\bar{D}$-pair production \cite{Maciula:2013wg}. As also
discussed in Refs.~\cite{Maciula:2019izq,Maciula:2016kkx}, the KMR model
effectively includes extra emission of hard partons (gluons) from the
uPDF that corresponds to higher-order contributions and leads therefore
to results well consistent with collinear NLO approach.
In order to investigate the impact of new dynamical effects -- beyond
those included in the DGLAP equation -- that are expected to be present
in the small-$x$ regime, we will also estimate the charm cross section
using as input the uPDF's obtained in Ref.~\cite{KutakSapeta} as
a solution of the Balitsky-Kovchegov equation \cite{bk} modified to
include the sub-leading corrections in $\ln (1/x)$ which are given by a
kinematical constraint, DGLAP $P_{gg}$ splitting function and the
running of the strong coupling (for a detailed derivation see
Ref.~\cite{KutakStasto}). Such an approach includes the corrections
associated with the BFKL equation, in an unified way with the DGLAP one, 
as well the nonlinear term, which takes into account unitarity
corrections.  
In Ref.~\cite{KutakSapeta} the authors performed a fit to the
combined HERA data and provided the solutions with and without the
inclusion of the nonlinear term. In the next section, we will use 
these solutions as input in our calculations and the corresponding
predictions will be denoted KS nonlinear and KS linear, respectively. 
For a comparison between predictions for the KMR, KS linear and KS
nonlinear ${\cal{F}}_{g^*} (x_2, k_t^2, \mu^2)$ we refer the interested reader
to Fig. 7 in Ref.~\cite{antoni_rafal_jhep}.

\section{Results}
\label{res}

\begin{figure}[t]
\begin{center}
\begin{tabular}{cc}
\includegraphics[width=0.5\textwidth]{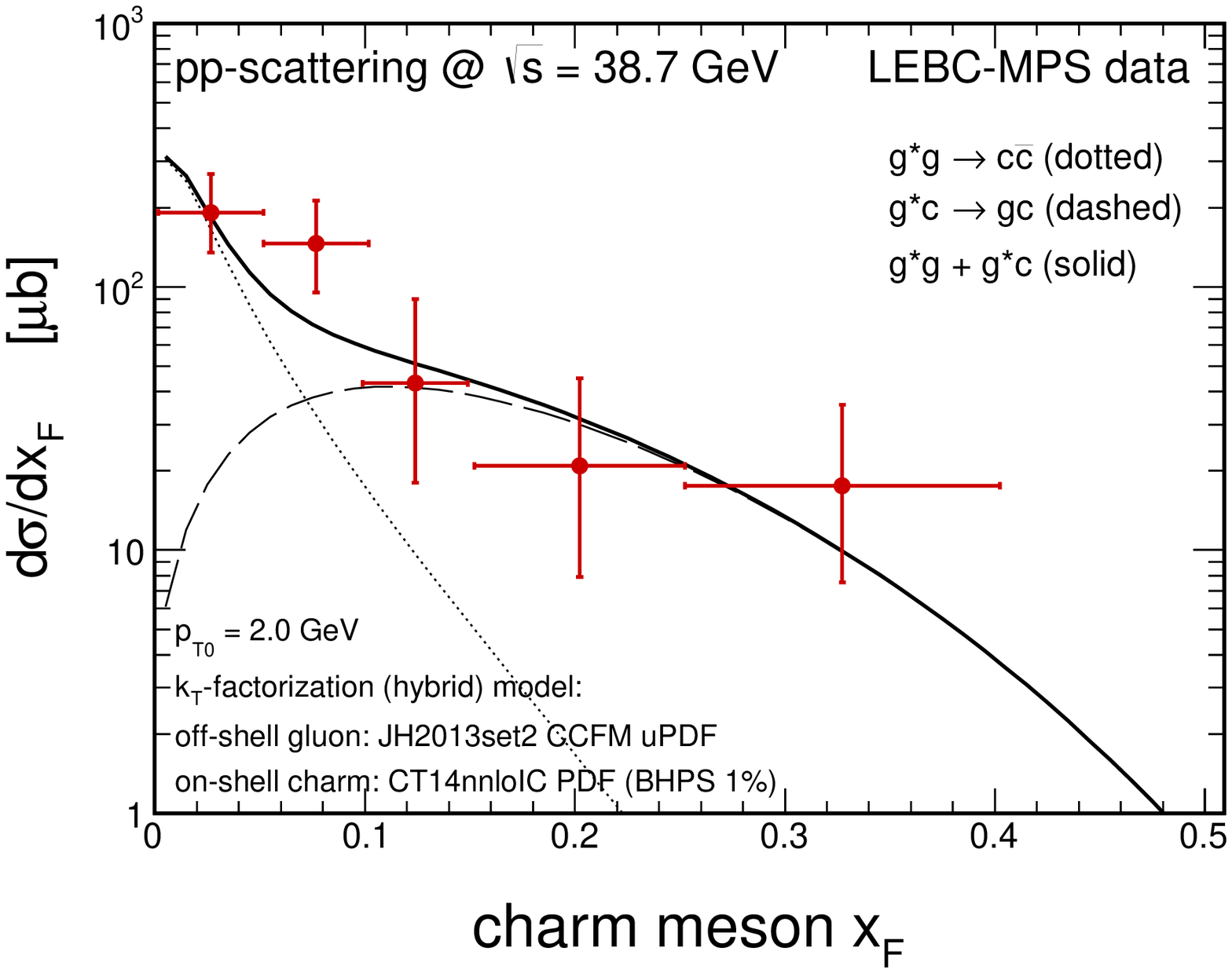} &  
\includegraphics[width=0.5\textwidth]{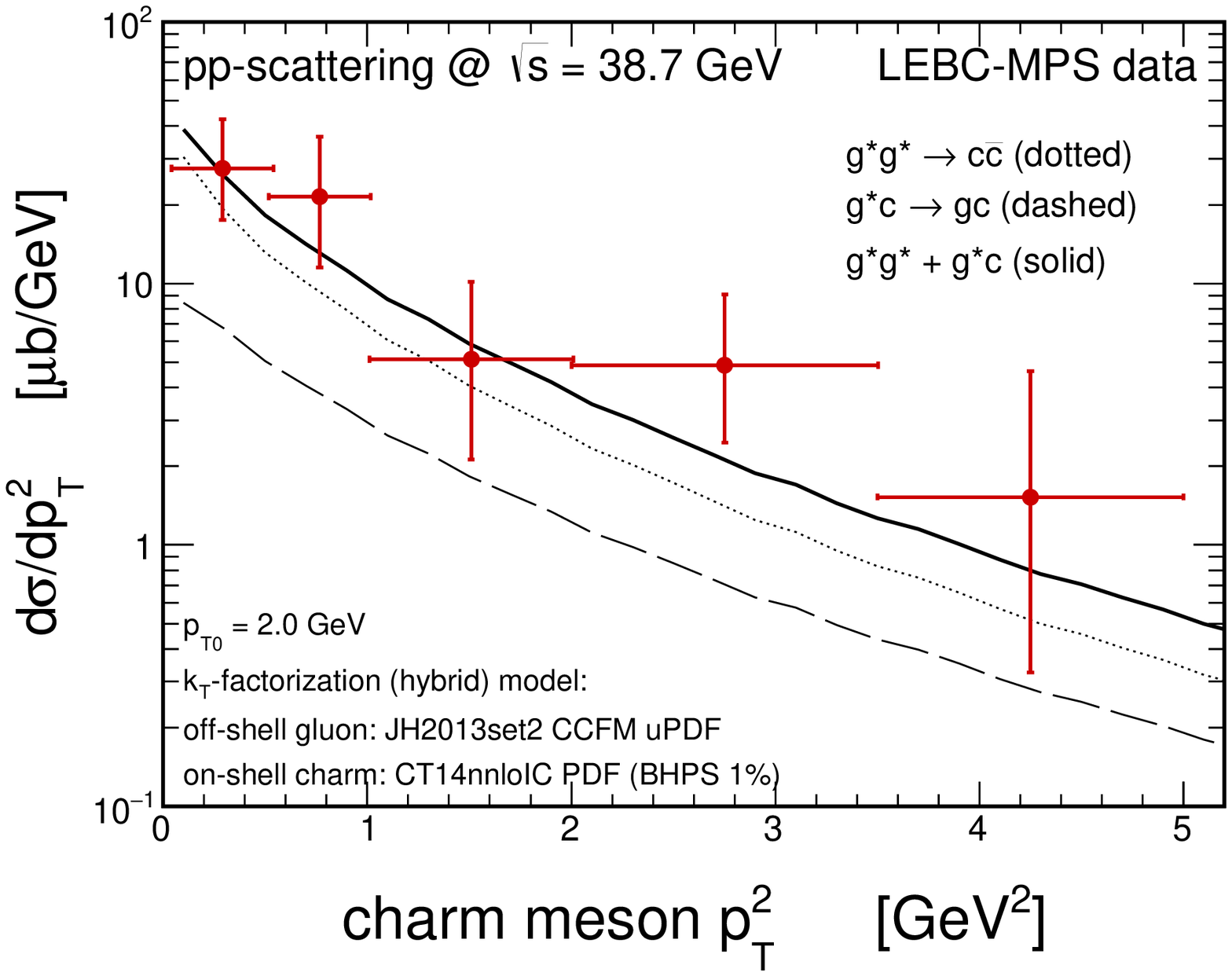} \\
(a) & (b)
\end{tabular}

\end{center}
\caption{Predictions of the hybrid model for (a) the Feynman $x_F$ -
   and (b) the transverse momentum distributions of charm particles
   produced in $pp$ collisions at $\sqrt{s} = 39$ GeV.
   Presented here data are from Ref. \cite{Ammar:1988ta}.}
\label{Fig:lebc-data}
\end{figure} 

In what follows we will present our predictions for the prompt
atmospheric neutrino flux derived using the $Z$-moment method. 
The effective hadronic interaction lengths $\Lambda_i$ and the $Z_{pp}$,
$Z_{HH}$ and $Z_{H\nu}$-moments will be estimated following
Ref.~\cite{anna1}. 
On the other hand, the $Z_{pH}$-moment will be calculated using as
input the $x_F$-distribution for the charm production derived in 
the hybrid approach with the ingredients discussed in the previous
section.  Moreover, the prompt $\nu_{\mu}$ flux will be evaluated
considering the description of the primary 
spectrum proposed by Gaisser in Ref.~\cite{gaisser}, denoted as 
H3a spectrum, which assumes that it is given by a composition
of 3 populations and 5 representative nuclei, with the set of parameters
determined by a global fit of the cosmic ray data.

\begin{figure}[t]
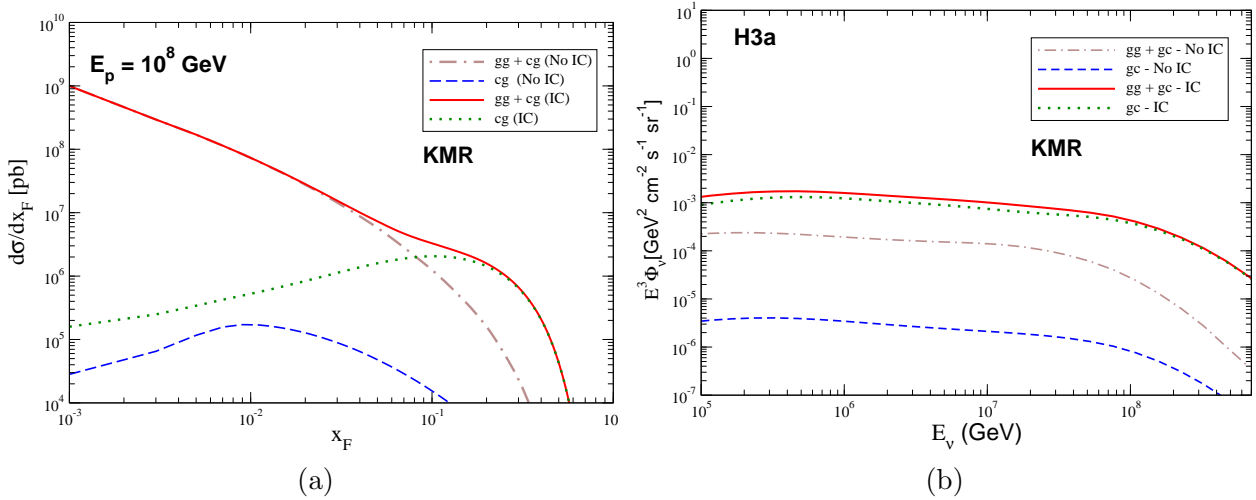

\begin{center}
\begin{tabular}{cc}
\includegraphics[width=0.5\textwidth]{dsigdxf_icxnoic_KMR_pt2.eps} &  
\includegraphics[width=0.5\textwidth]{fluxE3_KMR_pt2.eps} \\
(a) & (b)
\end{tabular}

\end{center}
\caption{Predictions of the hybrid model for (a) the Feynman $x_F$ -
  distributions for charm particles and (b) the prompt neutrino flux
  (rescaled by $E_{\nu}^3$),  calculated using the KMR model for the
  uPDF. The predictions with and without the presence of an intrinsic
  charm (here $p_{t0}$ = 2 GeV was used) are presented separately.
The H3a prametrization of the cosmic ray flux is used in this calculation.}
\label{Fig:kmr}
\end{figure} 

\begin{figure}[t]
\begin{center}
\includegraphics[width=0.65\textwidth]{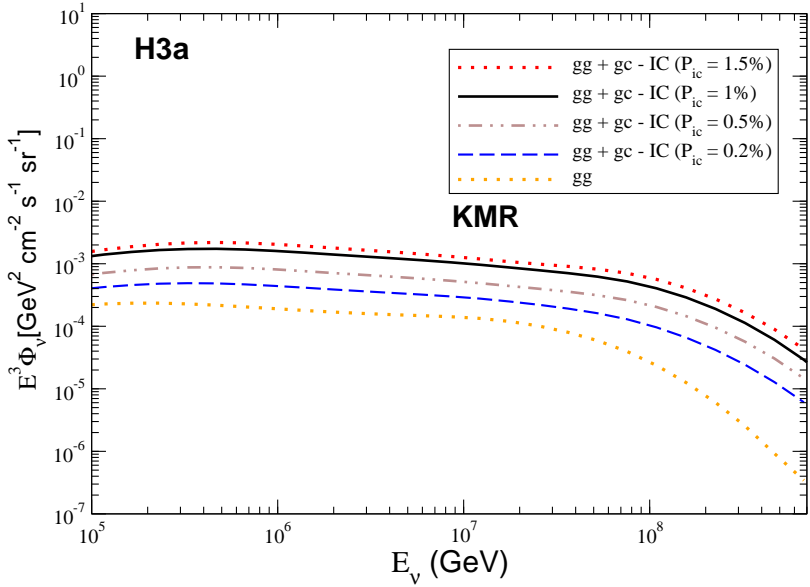} 
\end{center}
\caption{Predictions of the hybrid model for the  the prompt neutrino
  flux (rescaled by $E_{\nu}^3$),  calculated using the KMR model for
  the uPDF. The IC contribution was obtained with $p_{t0}$ = 2 GeV
    and assuming different values for the probability to find an
  intrinsic charm.
The H3a prametrization of the cosmic ray flux is used in this calculation.}
\label{Fig:pic}
\end{figure} 

As discussed in the previous Section, the predictions for the 
$cg \rightarrow cg$ mechanism are calculated assuming that the partons
in the final state are massless, which implies introduction of a
cutoff $p_{t0}$ to regularize the singularity in the partonic cross 
section (see \cite{antoni_rafal_jhep}). In order to constrain this parameter, 
we will initially consider the LEBC - MPC data \cite{Ammar:1988ta} 
for the $D$ meson production in $pp$ collisions at $\sqrt{s} = 39$ GeV. 
In Fig. \ref{Fig:lebc-data} we present our predictions for 
the $x_F$ and $p_T$ distributions of the charm meson,  obtained using 
the CT14nnloIC parametrization for $P_{ic} = 1\%$ in the calculation of 
the $cg \rightarrow cg$ mechanism. The results for the $x_F$
distribution indicate that the inclusion of the $cg^* \rightarrow cg$ 
mechanism is needed in order to describe the data. Moreover, 
the $p_T$ distribution is also  well described. 
Both results point out that a value of $p_{t0} = 2.0$ GeV is a good 
choice for the cutoff, which can be considered conservative, since 
smaller values imply a larger amount for the contribution of 
the $cg \rightarrow cg$ mechanism. Such choice is also justified 
by the recent analysis performed in Ref. \cite{Maciula:2021orz}, 
where a comprehensive study of the impact of an intrinsic charm 
component on the D meson production in $pHe$ fixed - target collisions 
at the LHCb was performed. The results presented in 
Ref. \cite{Maciula:2021orz} indicate that the LHCb data can be 
well described assuming  $p_{t0} = 2.0$ GeV for a probability of 
1\% of finding a charm quark-antiquark pair in the proton wave function.

 In Fig.~\ref{Fig:kmr} (a), we present our predictions for the
Feynman $x_F$ distribution of charm particles produced in $pp$
collisions at the atmosphere, considering an incident proton with an
energy of $E_p = 10^8$ GeV and the KMR model for the uPDF. 
Similar conclusions are derived using the KS linear and KS nonlinear
uPDFs. We present separately the contribution associated with the $cg
\rightarrow cg$ mechanism and the sum of the two mechanisms, denoted  
by ``cg'' and  ``gg + cg'', respectively. 
Moreover, we compare the IC predictions, obtained using the CT14nnloIC
parametrization for $P_{ic} = 1\%$, with those obtained disregarding 
the presence of the intrinsic component (denoted No IC hereafter). 
One has that for small $x_F (\equiv  x_1 - x_2)$, the charm production
is dominated by the $gg \rightarrow c \bar{c}$ mechanism, which is
expected since for  $x_F   \approx 0$ and high energies both
longitudinal momentum fractions $x_i$ are very small and the proton
structure is dominated by gluons. For the No IC case, the  contribution
of the $cg \rightarrow cg$ mechanism is smaller than the gluon fusion one
for all values of $x_F$. In contrast, when intrinsic charm is included,
the behavior of the distribution in the intermediate $x_F$ range ($0.06
\le x_F \le 0.6$) is strongly modified. Such a behaviour is expected,
since for this kinematical range, the charm production depends on
the description of the partonic content of the incident proton at large
values of the Bjorken $x$ variable. 
As discussed in the previous section, the main impact of the presence of
an intrinsic charm is  that the charm distribution is enhanced at large
$x$  ($> 0.1$), becoming larger than the gluon distribution. As a
consequence, the presence of an intrinsic charm implies that the Feynman
$x_F$-distribution for large $x_F$ is dominated by the $cg \rightarrow cg$
mechanism. The impact on the predictions for the prompt neutrino flux is
presented in Fig.~\ref{Fig:kmr} (b). As expected from the analysis
performed in Ref.~\cite{Goncalves:2017lvq}, where we find that the
dominant contribution to the neutrino flux comes typically from $x_F$ in
the region $0.2<x_F<0.5$, one has that the  flux is enhanced  by one
order of magnitude when intrinsic charm is present. In agreement with
the results presented in Fig.~\ref{Fig:kmr} (a), the contribution of the
$cg \rightarrow cg$ mechanism is negligible for the No IC case. 
However, it becomes dominant in the IC case, with the normalization of
the prompt flux dependent on the amount of IC present in the
projectile proton, as demonstrated in Fig.~\ref{Fig:pic}, where we
compare the prediction derived assuming $P_{ic} = 1\%$, which is the
assumption present in the CT14nnloIC parametrization, with the results
obtained assuming different values for this probability in the
calculation of the $x_F$ distribution for the $cg \rightarrow cg$
mechanism.  As expected from Eqs. (\ref{eq:flux}), (\ref{eq:zpc}) and
(\ref{eq:dif2}), 
our results indicate that $\phi_{\nu}$ is linearly dependent on $P_{ic}$
and, therefore, a precise determination of the prompt neutrino flux can
be used, in principle, to constrain the amount of IC 
in the proton (see below).

\begin{figure}[t]
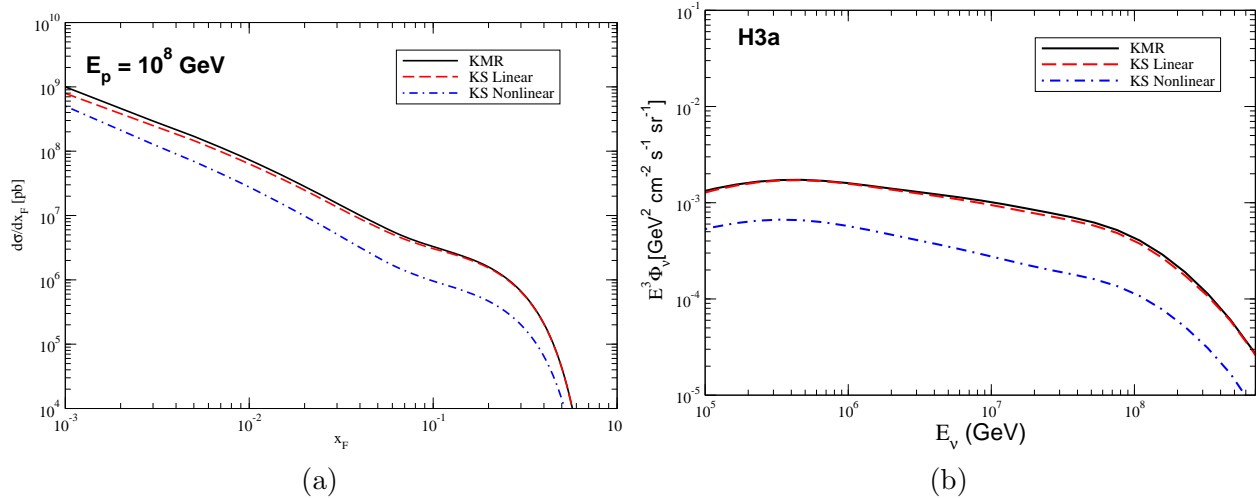

\begin{center}
\begin{tabular}{cc}
\includegraphics[width=0.5\textwidth]{dsigdxf_comp_models_pt2.eps} &  
\includegraphics[width=0.5\textwidth]{fluxE3_comp_models_pt2.eps} \\
(a) & (b)
\end{tabular}

\end{center}
\caption{Predictions of the hybrid model for the (a) Feynman $x_F$ -
  distributions for charm particles and (b) the prompt neutrino flux
  (rescaled by $E_{\nu}^3$),  
derived assuming different models for the uPDF.
The H3a prametrization of the cosmic ray flux is used in this calculation.}
\label{Fig:models}
\end{figure} 

The charm production at large $x_F$ is also dependent on the small-$x$
content of the target proton, which is dominated by  gluons.
The dependence of our results on the model assumed to describe the
unintegrated gluon distribution is analyzed in Fig.~\ref{Fig:models},
where we present the predictions for the $x_F$ distribution and for the
prompt neutrino flux derived assuming the KMR, KS linear and KS
nonlinear models as input in our calculations. For this analysis, we
only present the sum of the two mechanisms for charm production and the IC
predictions. 
One has that KMR and KS linear predictions for the $x_F$
distribution are similar, with the KMR one being slightly larger for
small $x_F$. On the other hand, the KS nonlinear is a factor $\approx 3$
smaller for $x_F = 0.2$. Such a result demonstrates that the inclusion of
the BFKL effects in modelling ${\cal{F}}_{g^*}$ has a small
effect on the behaviour of the distribution for large $x_F$. In
contrast, the inclusion of the nonlinear (saturation) effects strongly
modifies the magnitude of the distribution.
  A similar conclusion is 
derived from the analysis of Fig.~\ref{Fig:models}(b), where we
present our predictions for the prompt neutrino flux. One important
aspect is that the saturation effects imply a suppression 
of the flux in the kinematical range probed by the IceCube ($E_{\nu} \lesssim 10^7$ GeV).

\begin{figure}[t]
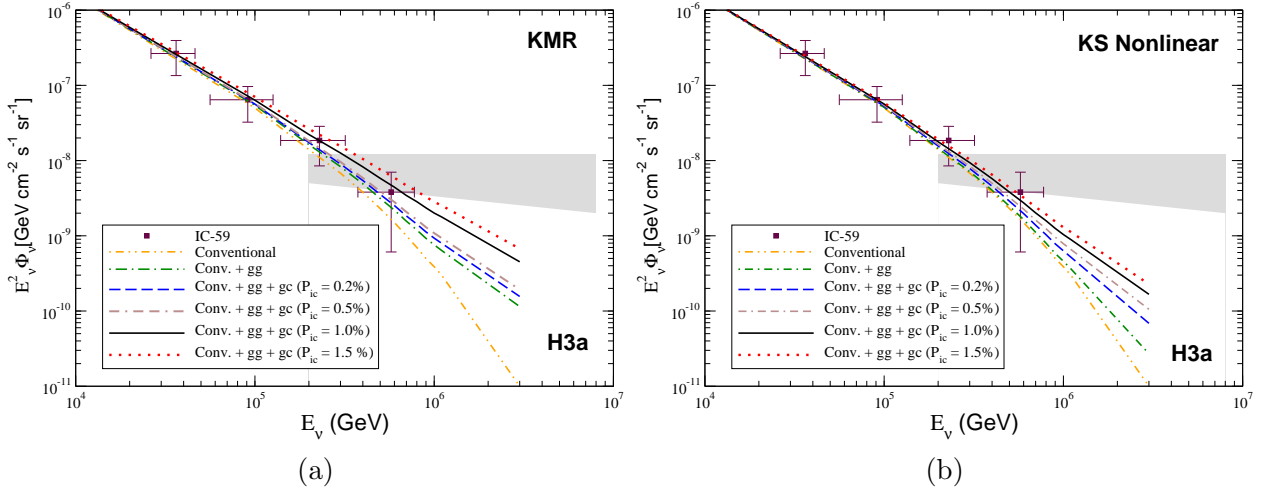

\begin{center}
\begin{tabular}{cc}
\includegraphics[width=0.5\textwidth]{fluxE2_points_H3a_KMR_pt2.eps} &  
\includegraphics[width=0.5\textwidth]{fluxE2_points_H3a_KSNonlinear2_pt2.eps} \\
(a) & (b)
\end{tabular}

\end{center}
\caption{Comparison between our predictions and the experimental IceCube
  data \cite{ice1} for the atmospheric $\nu_{\mu}$ flux for 
  (a) KMR and (b) KS nonlinear uPDFs. The IC contribution was obtained
  for $p_{t0}$ = 2 GeV as discussed in the main text.
The H3a prametrization of the cosmic ray flux is used in this
calculation.
The shaded band represents the
results from Ref.~\cite{Aartsen:2016xlq} for the astrophysical neutrino
flux.}
\label{Fig:fluxe2}
\end{figure} 

Our results indicate that the presence of the intrinsic charm implies
enhancement of the prompt $\nu_{\mu}$ flux, while the saturation
effects suppress it for high energies. Another important aspect is that
the impact of the $cg \rightarrow cg$ mechanism depends on the magnitude
of $P_{ic}$. One important question is whether the current or future
experimental IceCube data can be used to probe the presence of these
effects and constrain the probability to find an IC on the proton
structure, i.e. whether those data could help to improve our understanding
of the strong interactions theory.  
In recent years the IceCube Collaboration measured the energy
spectrum of atmospheric neutrino flux with larger precision in an
extended energy range \cite{ice1,Aartsen:2016xlq} and more data are
expected in the forthcoming years
\cite{Aartsen:2019kpk,Aartsen:2020fgd}. 
Such measurements are a challenge due to steeper falling behaviour 
expected for the atmospheric flux in comparison to that associated with
 astrophysical neutrinos. 
Different methods have been proposed to disentangle these 
two contributions with promising results (see
e.g. Ref.~\cite{Aartsen:2019kpk}).  
Therefore, the posed question is valid, relevant and timely.

The IceCube apparatus can measure directions of neutrinos/antineutrinos
\cite{review_neutrinos}. The IceCube experimental data discussed below 
is selected
taking into account only such $\nu_{\mu}$ neutrinos that passed through
the Earth (see Fig.~\ref{Fig:IceCube}).
In Fig.~\ref{Fig:fluxe2} we present our results for the atmospheric
$\nu_{\mu}$ flux, scaled by a factor $E_{\nu}^2$, which is the sum of
the conventional and prompt contributions. The predictions were obtained
considering different models for the uPDFs and distinct values for
$P_{ic}$ in the calculation of the prompt contribution. Moreover, for
the conventional  atmospheric neutrino flux we assume the result derived
in Ref.~\cite{Honda:2006qj}. The resulting predictions are compared with
the IceCube data obtained in Ref.~\cite{ice1} for the zenith-averaged flux of atmospheric neutrinos. For completeness, the
results from Ref.~\cite{Aartsen:2016xlq} for the astrophysical neutrino
flux are represented by the grey band. 
 One has that the prompt contribution enhances the flux at large
 neutrino energies, with the enhancement being strongly dependent on the
 magnitude of the $cg \rightarrow cg$ mechanism and the uPDF considered as
 input in the calculations. If this mechanism is disregarded, the results
 represented by  ``Conv. + gg'' in the figures indicate that the impact
 of the prompt flux is small in the current kinematical range probed by
 IceCube. In particular, it is negligible when the saturation effects are
 taken into account [see Fig.~\ref{Fig:fluxe2} (b)]. 
On the other hand, the inclusion of the $cg \rightarrow cg$ mechanism
implies a large enhancement of the prompt flux at large $E_{\nu}$, with
the associated magnitude being strongly dependent on the value of
$P_{ic}$. Our results for the KMR uPDF, presented in Fig.~\ref{Fig:fluxe2} (a), indicate that a value of $P_{ic}$ larger than $1.5 \%$ implies a prediction
for neutrino flux that overestimate the IceCube data at high energies. 
We have verified that a similar result is obtained for the KS linear 
uPDF (not shown explicitly). Therefore, the results derived assuming 
that the QCD dynamics is described by  linear evolution equations, 
which disregard the saturation effects, indicate that in order 
to describe the current IceCube data we
should have  $P_{ic} \lesssim 1.5 \%$. Surely, future data can be more
restrictive in the acceptable range of values for $P_{ic}$.  In
contrast, the results presented in Fig.~\ref{Fig:fluxe2} (b) suggest 
the presence of  saturation effects with  $P_{ic} = 1.5\%$ is not
discarded by the current IceCube data.
It is important to emphasize that the values of the $P_{ic}$ probabilities suggested above could be slightly
decreased if a smaller value of the $p_{t0}$ parameter was used 
in the numerical calculations of the $cg^* \to cg$ cross section.

However, from these results we can conclude that currently we have two acceptable
solutions when the $cg \rightarrow cg$ mechanism is included in the
analysis: (a) the QCD dynamics is described by a linear evolution
equation and the amount of IC in the proton wave function is similar to that predicted by 
the CT14nnloIC parameterization; or (b) the amount of IC is larger than that 
described by the CT14nnloIC parameterization and the saturation effects
are needed to describe the charm production at very forward
rapidities. One has that if the amount of IC is constrained in hadronic
colliders, the IceCube data for the atmospheric 
neutrino flux can be considered as a probe of the QCD dynamics at high
energies. Inversely, if the saturation effects are probed in hadronic 
colliders, the IceCube data can be used to constrain the amount of the
IC. Such results demonstrate 
synergy between IceCube and the LHC, and strongly motivate new experimental
and theoretical analyses in the future.

\section{Summary}
\label{conc}

One of the main goals of the IceCube observatory is the study
of astrophysical neutrinos. In order to separate the associated 
component, it is fundamental to have theoretical control of 
the background related to the atmospheric neutrino flux, where 
the neutrinos are generated from the decay of particles produced 
in high energy interactions between the Cosmic Rays and the atmosphere. 
In particular, the contribution of the prompt neutrino flux is still 
a theme of intense debate, since its magnitude for 
the IceCube Observatory and future neutrino telescopes depends on 
our knowledge about the QCD dynamics at high energies and on 
the large-$x$ $c \bar c$ partonic content of hadrons. 
In this paper, we have investigated the impact of the intrinsic charm 
component in the hadron wave function, 
which carries a large fraction of the hadron momentum, 
and from saturation effects, associated with nonlinear corrections 
in the QCD evolution, in the prompt neutrino flux.  Our results 
has indicated that
the inclusion of the $cg \rightarrow cg$ mechanism has a strong effect 
on the prompt neutrino flux. In particular, when the IC component 
is present, such a mechanism determines the energy dependence of 
the flux at high energies, with the normalization dependent on 
the value assumed for the probability to find the IC in 
the proton wave function.  
Furthermore, we find that the saturation effects suppress the
prompt flux in the kinematical range probed by the IceCube. The
comparison of our predictions with the current IceCube experimental data
has indicated that for a linear QCD dynamics, $P_{ic}$ can be of the
order of the value  assumed  by the CT14nnlo parametrization. In
contrast, a somewhat larger value is still acceptable when a nonlinear QCD
dynamics is included.
Consequently, in order to
disentangle these two possibilities, it is mandatory to have a better
theoretical and experimental control of the prompt neutrino flux at
IceCube and of the charm production at the LHC. Such a result strongly
motivates the analysis of other processes that allow us to probe the
presence of the intrinsic charm and contrain the description of the QCD
dynamics at high energies. One of such alternatives is 
the analysis of the $D$-meson and $\nu_{\mu}$ neutrino production 
at FASER \cite{Feng:2017uoz} taking into account both effects, which 
we intend to study in a forthcoming publication.
  

\section*{Acknowledgments}
VPG was  partially financed by the Brazilian funding
agencies CNPq,   FAPERGS and  INCT-FNA (process number 
464898/2014-5).
components in mesons.
This study was also partially supported by the Polish National Science Center
grant UMO-2018/31/B/ST2/03537 and by the Center for Innovation and
Transfer of Natural Sciences and Engineering Knowledge in Rzesz{\'o}w.




\begin{thebibliography}{99}

\bibitem{pdg}
P.~A.~Zyla \textit{et al.} [Particle Data Group],
PTEP \textbf{2020}, no.8, 083C01 (2020).

\bibitem{ice1}
M.~G.~Aartsen \textit{et al.} [IceCube],
Eur. Phys. J. C \textbf{75}, no.3, 116 (2015).

\bibitem{Aartsen:2016xlq} 
  M.~G.~Aartsen {\it et al.} [IceCube Collaboration],
  Astrophys.\ J.\  {\bf 833}, 3 (2016).

\bibitem{Aartsen:2017nbu}
M.~G.~Aartsen \textit{et al.} [IceCube],
Eur. Phys. J. C \textbf{77}, no.10, 692 (2017).  
  
 
\bibitem{Aaij:2013mga}
R.~Aaij \textit{et al.} [LHCb],
Nucl. Phys. B \textbf{871}, 1-20 (2013).

\bibitem{Aaij:2015bpa}
R.~Aaij \textit{et al.} [LHCb],
JHEP \textbf{03}, 159 (2016)
[erratum: JHEP \textbf{09}, 013 (2016); erratum: JHEP \textbf{05}, 074 (2017).]

\bibitem{Acharya:2017jgo}
S.~Acharya \textit{et al.} [ALICE],
Eur. Phys. J. C \textbf{77}, no.8, 550 (2017).

\bibitem{Acharya:2019mgn}
S.~Acharya \textit{et al.} [ALICE],
Eur. Phys. J. C \textbf{79}, no.5, 388 (2019).
  
\bibitem{Goncalves:2017lvq}
V.~P.~Goncalves, R.~Maciu\l{}a, R.~Pasechnik and A.~Szczurek,
Phys. Rev. D \textbf{96}, no.9, 094026 (2017).

\bibitem{review_neutrinos}   M.~Ahlers, K.~Helbing and C.~P\'erez de los Heros,
Eur. Phys. J. C \textbf{78}, no.11, 924 (2018); M.~Ahlers and F.~Halzen,
Prog. Part. Nucl. Phys. \textbf{102}, 73-88 (2018).

\bibitem{sigl} 
  M.~V.~Garzelli, S.~Moch and G.~Sigl,
  JHEP {\bf 1510}, 115 (2015).

\bibitem{anna1}  
  A.~Bhattacharya, R.~Enberg, M.~H.~Reno, I.~Sarcevic and A.~Stasto,
  JHEP {\bf 1506}, 110 (2015).

\bibitem{gauld}  
  R.~Gauld, J.~Rojo, L.~Rottoli and J.~Talbert,
  JHEP {\bf 1511}, 009 (2015); R.~Gauld, J.~Rojo, L.~Rottoli, S.~Sarkar and J.~Talbert,
  JHEP {\bf 1602}, 130 (2016).

\bibitem{halzen}  
   F.~Halzen and L.~Wille,
  Phys.\ Rev.\ D {\bf 94}, 014014 (2016).

\bibitem{anna2} 
  A.~Bhattacharya, R.~Enberg, Y.~S.~Jeong, C.~S.~Kim, M.~H.~Reno, I.~Sarcevic and A.~Stasto,
  JHEP {\bf 1611}, 167 (2016).
  
\bibitem{laha}  
     R.~Laha and S.~J.~Brodsky,
  Phys.\ Rev.\ D {\bf 96}, 123002 (2017).
    
\bibitem{prosa}  
     M.~V.~Garzelli {\it et al.} [PROSA Collaboration],
  JHEP {\bf 1705}, 004 (2017);   M.~Benzke, M.~V.~Garzelli, B.~Kniehl, G.~Kramer, S.~Moch and G.~Sigl,
  JHEP {\bf 1712}, 021 (2017); O.~Zenaiev \textit{et al.} [PROSA],
JHEP \textbf{04}, 118 (2020).
  

 \bibitem{prdandre}
    A.~V.~Giannini, V.~P.~Goncalves and F.~S.~Navarra,
  Phys.\ Rev.\ D {\bf 98}, no. 1, 014012 (2018).


\bibitem{Goncalves:2018zzf}
V.~P.~Goncalves, R.~Maciu\l{}a and A.~Szczurek,
Phys. Lett. B \textbf{794}, 29-35 (2019).  

\bibitem{bhps}        S.~J.~Brodsky, P.~Hoyer, C.~Peterson and N.~Sakai,
                      Phys.\ Lett.\  B {\bf 93}, 451 (1980).



\bibitem{pnndb}       S.~Paiva, M.~Nielsen, F.~S.~Navarra, F.~O.~Duraes and L.~L.~Barz, 
                      Mod.\ Phys.\ Lett.\  A {\bf 13}, 2715 (1998);
                      F.~S.~Navarra, M.~Nielsen, C.~A.~A.~Nunes and M.~Teixeira, 
                      Phys.\ Rev.\  D {\bf 54}, 842 (1996).

\bibitem{wally99}     F.~M.~Steffens, W.~Melnitchouk and A.~W.~Thomas,  Eur.\ Phys.\ J.\ C {\bf 11}, 673 (1999).


\bibitem{nikwolf}
N.~N.~Nikolaev and W.~Schafer,
Phys. Rev. D \textbf{71}, 014023 (2005).

\bibitem{raju}
H.~Fujii, F.~Gelis and R.~Venugopalan,
Phys. Rev. Lett. \textbf{95}, 162002 (2005).


\bibitem{Dominguez:2010xd}
F.~Dominguez, B.~W.~Xiao and F.~Yuan,
Phys. Rev. Lett. \textbf{106}, 022301 (2011).

\bibitem{Kotko:2015ura}
P.~Kotko, K.~Kutak, C.~Marquet, E.~Petreska, S.~Sapeta and A.~van Hameren,
JHEP \textbf{09}, 106 (2015).


\bibitem{cgc}         F.~Gelis, E.~Iancu, J.~Jalilian-Marian and R.~Venugopalan, 
                      Ann.\ Rev.\ Nucl.\ Part.\ Sci.\  {\bf 60}, 463 (2010);
                      E.~Iancu and R.~Venugopalan, arXiv:hep-ph/0303204; H.~Weigert,  
                      Prog.\ Part.\ Nucl.\ Phys.\ {\bf 55}, 461 (2005); 
                      J.~Jalilian-Marian and Y.~V.~Kovchegov,
                      Prog.\ Part.\ Nucl.\ Phys.\  {\bf 56}, 104 (2006);   
                      J.~L.~Albacete and C.~Marquet,  Prog.\ Part.\ Nucl.\ Phys.\  {\bf 76}, 1 (2014).    


\bibitem{Maciula:2017wov}
R.~Maciu{\l}a and A.~Szczurek,
Phys. Rev. D \textbf{97}, no.7, 074001 (2018).

\bibitem{antoni_rafal_jhep}
R.~Maciu\l{}a and A.~Szczurek,
JHEP \textbf{10}, 135 (2020).

\bibitem{ingelman} 
  P.~Gondolo, G.~Ingelman and M.~Thunman,
  Astropart.\ Phys.\  {\bf 5}, 309 (1996).
  
  
  \bibitem{Bugaev:1998bi}
E.~V.~Bugaev, A.~Misaki, V.~A.~Naumov, T.~S.~Sinegovskaya, S.~I.~Sinegovsky and N.~Takahashi,
Phys. Rev. D \textbf{58}, 054001 (1998)




  \bibitem{Honda:2006qj} 
  M.~Honda, T.~Kajita, K.~Kasahara, S.~Midorikawa and T.~Sanuki,
  Phys.\ Rev.\ D {\bf 75}, 043006 (2007).

\bibitem{Ostapchenko:2005nj}
  S.~Ostapchenko,
  Phys.\ Rev.\ D {\bf 74}, no. 1, 014026 (2006).
  
  \bibitem{vicandre}
A.~V.~Giannini and V.~P.~Gon\c{c}alves,
Eur. Phys. J. C \textbf{79}, no.2, 158 (2019).

\bibitem{Brodsky:2020zdq}
S.~J.~Brodsky, G.~I.~Lykasov, A.~V.~Lipatov and J.~Smiesko,
Prog. Part. Nucl. Phys. \textbf{114}, 103802 (2020).


\bibitem{nnpdf}
R.~D.~Ball \textit{et al.} [NNPDF],
Eur. Phys. J. C \textbf{76}, no.11, 647 (2016).


\bibitem{ct14}
T.~J.~Hou, S.~Dulat, J.~Gao, M.~Guzzi, J.~Huston, P.~Nadolsky, C.~Schmidt, J.~Winter, K.~Xie and C.~P.~Yuan,
JHEP \textbf{02}, 059 (2018).


\bibitem{wally16}     P.~Jimenez-Delgado, T.~J.~Hobbs, J.~T.~Londergan and W.~Melnitchouk,
                      Phys.\ Rev.\ Lett.\  {\bf 116}, 019102 (2016).

\bibitem{brod_letter} S.~J.~Brodsky and S.~Gardner,  Phys.\ Rev.\ Lett.\  {\bf 116}, 019101 (2016).

 \bibitem{bk}  I.~Balitsky,
Nucl. Phys. B \textbf{463}, 99-160 (1996); Y.~V.~Kovchegov,
Phys. Rev. D \textbf{60}, 034008 (1999).

\bibitem{kmr}
G.~Watt, A.~D.~Martin and M.~G.~Ryskin,
Eur. Phys. J. C \textbf{31}, 73-89 (2003).

\bibitem{Maciula:2013wg}
R.~Maciu{\l}a and A.~Szczurek,
Phys. Rev. D \textbf{87}, no.9, 094022 (2013).

\bibitem{Maciula:2019izq}
R.~Maciu{\l}a and A.~Szczurek,
Phys. Rev. D \textbf{100}, no.5, 054001 (2019).

\bibitem{Maciula:2016kkx}
R.~Maciu{\l}a and A.~Szczurek,
Phys. Rev. D \textbf{94}, no.11, 114037 (2016).


\bibitem{KutakSapeta}
K.~Kutak and S.~Sapeta,
Phys. Rev. D \textbf{86}, 094043 (2012).


\bibitem{KutakStasto}
K.~Kutak and A.~M.~Stasto,
Eur. Phys. J. C \textbf{41}, 343-351 (2005).

\bibitem{Maciula:2019iak}
R.~Maciu\l{}a and A.~Szczurek,
J. Phys. G: Nucl. Part. Phys. 47, 035001 (2020).

\bibitem{Peterson:1982ak}
C.~Peterson, D.~Schlatter, I.~Schmitt and P.~M.~Zerwas,
Phys. Rev. D \textbf{27}, 105 (1983).

\bibitem{gaisser}     T.~K.~Gaisser,
                       Astropart.\ Phys.\  {\bf 35}, 801 (2012).


\bibitem{Ammar:1988ta}
R.~Ammar, R.~C.~Ball, S.~Banerjee, P.~C.~Bhat, P.~Bosetti, C.~Bromberg, G.~E.~Canough, T.~Coffin, T.~O.~Dershem and R.~L.~Dixon, \textit{et al.}
Phys. Rev. Lett. \textbf{61}, 2185-2188 (1988)

\bibitem{Maciula:2021orz}
R.~Maciula and A.~Szczurek,
[arXiv:2105.09370 [hep-ph]].

\bibitem{Aartsen:2019kpk}
M.~G.~Aartsen \textit{et al.} [IceCube],
[arXiv:1907.11699 [astro-ph.HE]].

\bibitem{Aartsen:2020fgd}
M.~G.~Aartsen \textit{et al.} [IceCube Gen2],
[arXiv:2008.04323 [astro-ph.HE]].

 
\bibitem{Feng:2017uoz}
J.~L.~Feng, I.~Galon, F.~Kling and S.~Trojanowski,
Phys. Rev. D \textbf{97}, no.3, 035001 (2018).


\end{thebibliography}
\end{document}